\documentclass[twocolumn,showpacs, prl]{revtex4-1}
\usepackage{graphicx}
\usepackage{float}
\usepackage{epstopdf}
\usepackage{parskip}
\usepackage{setspace}
\usepackage{amsmath}
\usepackage{color}
\usepackage{soul}
\usepackage{tabularx}
\usepackage{xspace}
\begin{document}

\title{On the Theory of Indirect Exchange in EuO}

\author{Shmuel Burg}
\email{burgshmuel@gmail.com}

\author{Vladimir Stukalov}
\email{vova.stukalov@gmail.com}

\author{Eugene Kogan}
\email{Eugene.Kogan@biu.ac.il}

\affiliation{Department of Physics, Bar-Ilan University, Ramat-Gan 52900,
Israel}
\date{\today}

\begin{abstract}
We present the calculations of the Curie temperature and magnetization of  doped EuO both in the absence and in the presence of external magnetic field. The calculations were performed both for the free electrons model and for the model with finite electron band width. Both models give similar results for the magnetization, close to Brillouin function.

\end{abstract}


\maketitle
\section{Introduction}
The perspectives of the development of spintronics has led to renewed interest in rear-earth oxide ferromagnetic semiconductors such as europium chalcogenides, in particular
EuO and EuS, which have a rock salt (fcc) structure, with a lattice constant of $5.14$ A for EuO and $5.96$ A for EuS, and whose magnetism arises from partially filled and highly localized $4f$ states. Stoichiometric EuO and EuS are regarded as  typical Heisenberg ferromagnets, with a Curie temperature ($T_C$)  69.8 K for EuO and 16.6 K for EuS respectively \cite {metfessel,young}. From the point of view of electronic properties these
 materials are semiconductors, with a  band gap  at room temperature of 1.12 eV for EuO and .6 eV for EuS.
For applications EuO looks especially promising \cite{mairoser}; it has the third
strongest saturation magnetization of all known ferromagnets \cite{mathias}, one of the largest magneto-optic Kerr effects \cite{ahn},
pronounced insulator-to-metal transitions \cite{shafer,kogan,sinjukow,kroha} as well as
colossal magnetoresistance effects \cite{shapira}.
For the use of these materials  in spintronics applications  it is desirable to increase $T_C$ as much as possible. This can be achieved, in particular, by doping EuO with the rear earth metals with valency 3 (Gd \cite{mairoser}, La
\cite{miyazaki}) or with oxygen vacancies. Introduction of carriers into the conduction band by doping leads to indirect exchange between the localized spins, thus  the Curie temperature depends upon the electron concentration.

The seminal calculations of the influence of the indirect exchange on magnetic properties of doped Europium oxides were performed by Mauger \cite{mauger}. In these calculations  the $s-f$ interaction is treated in the second order of perturbation theory,
very much similar to the RKKY theory \cite{ruderman}. However, the RKKY approach was substantially modified due to some specific features of magnetic semiconductors.
First, in  magnetic semiconductors  the electronic gas is not always degenerate, contrary to metals. Second, due to the fact that the atomic exchange is not small compared to Fermi energy, the temperature dependence of the indirect exchange
cannot be neglected. Thus the calculations were performed explicitly at a finite temperature $T$ and not at $T=0$ as usual. Also
the conduction band of a finite width was considered, contrary to the RKKY assumption of the free electrons dispersion law.
Mauger calculations still remain an important reference point for the experimentalists in the field, including recent experiments \cite{mairoser,Barbagallo}.

The theory of indirect exchange was substantially modified  by Nolting \cite{Nolting1}\cite{Nolting2}\cite{Nolting3}. The modified  theory involves 3 types of correlation functions, describing  itinerant electrons correlations,  local-spin correlations and mixed itinerant-electron-local-moment correlations.
This theory is able to describe magnetic semiconductors (EuO, EuS), diluted magnetic semiconductors ($Ga_{1-x}Mn_xAs$), magnetic metals (Gd, Dy,Tb) and CMR materials($La_{1-x}Ca_xMnO_3$). Using this model Nolting had performed calculations of density of states (DOS) and energy band structure for EuO in $T=0$ and finite temperature.\cite{Nolting4}\cite{Nolting5}

However, in the present work we decided to limit ourselves with the Mauger type calculations. Our aim was to check up to what extent the final results are influenced by the dispersion law of the itinerant electrons. We compared the results obtained in the framework of the free electrons model with those obtained in the framework of the model with the finite electron band width used by Mauger \cite{mauger}.
As an additional modification we performed the calculations
using Matsubara Green functions to  calculate indirect exchange at a finite temperature.
\section{Indirect exchange by Matsubara Green functions}
The Hamiltonian we start from is\newline
\begin{eqnarray}
H&=&-\frac{1}{2m}\sum_{\alpha}\int \psi_{\alpha}^{\dagger}({\bf r})\nabla^2\psi_{\alpha}({\bf r})d^3{\bf r}-\frac{1}{2}\sum_{ij} I({\bf R}_{ij}) {\bf S}_i{\bf S}_j\nonumber\\
&+&J_{df}\sum_i{\bf S}_i{\bf\sigma}({\bf R}_i),
\end{eqnarray}
where ${\bf \sigma}({\bf r})=\psi_{\alpha}^{\dagger}({\bf r})\sigma_{\alpha\beta}\psi_{\alpha}({\bf r})$.

We'll use relevant for EuO approximation of weak $s-f$ exchange coupling, and calculate the indirect exchange between localized spins appearing due to their interaction with conduction electrons in the leading order of perturbation theory. (For the intermediate coupling  regime see e.g.
\cite{kogan2,stier}.)
Though our aim is to obtain finite temperature results for the indirect exchange, we'll start from the derivation of the RKKY interaction for $T=0$
as presented in Ref. \cite{levitov}.
We can write down spin polarization in the Fermi gas
$\sigma^{i}({\bf r})=\left\langle\psi^{\dagger}_{\alpha}({\bf r})\sigma^i_{\alpha\beta}\psi_{\beta}({\bf r})\right\rangle$ using Green function\newline
\begin{eqnarray}
\sigma^i({\bf r})=-i{\rm Tr}\left(\hat{\sigma}^i\hat{G}\right)=\lim_{t'\to t+0,{\bf r}={\bf r}'}
\left(-i\sigma^i_{\alpha\beta}G_{\beta\alpha}({\bf r},t;{\bf r}',t')\right),\nonumber\\
\end{eqnarray}\newline
where the summation on indices $\alpha,\beta$ is implied.
Let assume that there is a localized (at ${\bf r}=0$) spin ${\bf S}$  interacting with the local spin density of conduction electrons\newline
\begin{eqnarray}
H_{int}=J_{df}\hat{\bf S}^i\hat\sigma^i({\bf r}=0),
\end{eqnarray}\newline
where $\hat\sigma^i=\psi^{\dagger}_{\alpha}({\bf r})\sigma^i_{\alpha\beta}\psi_{\beta}({\bf r})$. In the first order with respect to interaction\newline
\begin{eqnarray}
G^{(1)}(\epsilon, {\bf r}),{\bf r}')=J_{df}S^i\sigma^i_{\alpha\beta}G_0(\epsilon, {\bf r})G_0(\epsilon, -{\bf r}').
\end{eqnarray}\newline
Hence\newline
\begin{eqnarray}
\label{free}
\sigma^i({\bf r})=-2iJ_{df}S^i\int G_0^2(\epsilon,{\bf r})\frac{d\epsilon}{2\pi}.
\end{eqnarray}\newline
\section{Free electrons model}
For the free electrons model we can take\newline
\begin{eqnarray}
\label{sq0}
G_0(\epsilon,{\bf p})=\frac{1}{\epsilon+E_F-p^2/2m+i\delta{\rm sign}\epsilon},
\end{eqnarray}\newline
and hence\newline
\begin{eqnarray}
\label{sq}
G_0(\epsilon,{\bf r})=-\frac{m}{2\pi r}e^{i{\rm sign}\epsilon\kappa r},
\end{eqnarray}\newline
where $\kappa=\sqrt{2m(E_F+\epsilon+i\delta{\rm sign}\epsilon)}$ and calculate the integral (\ref{free}) to obtain\newline
\begin{eqnarray}
\label{free2}
\sigma^i({\bf r})=J_{df}S^i\frac{2mk_F^4}{\pi^3}\left(\frac{\cos x}{x^3}-\frac{\sin x}{x^4}\right),
\end{eqnarray}\newline
where $x=2k_Fr$.
Thus we obtain  the well known RKKY result\newline
\begin{eqnarray}
H_{ex}'=-\frac{1}{2}\sum_{ij} J_{eff}({\bf R}_{ij}) S_i^zS^z_j,
\end{eqnarray}\newline
where\newline
\begin{eqnarray}
\label{free7}
&&J_{eff}(R_{ij})=\frac{2 J_{df}^2}{\mu}\frac{(k_Fa_0/2)^6}{\pi^3}\\
&&\frac{\sin(2k_FR_{ij})-2k_FR_{ij}\cos(2k_FR_{ij})}{(2k_FR_{ij})^4},\nonumber
\end{eqnarray}\newline
where $\mu=k_F^2/2m$ and $\lambda=k_F\tau/m^*$.
To make calculations for finite temperature \cite{levitov}, first we have to calculate\newline
\begin{eqnarray}
G(i\omega_n,{\bf r})=\int\frac{d^3p}{(2\pi)^3}\frac{e^{i{\bf pr}}}{i\omega_n-\xi_{\bf p}+\frac{i}{2\tau}{\rm sign}\;\omega_n},
\end{eqnarray}\newline
where $\xi_p=\frac{p^2}{2m}-E_F$ and  $\omega_n=(2n+1)\pi T$.
(We have  taken into account scattering of conduction electrons \cite{mattis}; $\tau$ is the scattering time.)
Making fraction decomposition we obtain\newline
\begin{eqnarray}
&&G(i\omega_n,{\bf r})=\frac{m}{4\pi^2r}\int_{-\infty}^{\infty}\left(\frac{1}{\kappa-p}-\frac{1}{\kappa+p}\right)\sin pr dp\nonumber\\
&&=-\frac{m}{2\pi r}e^{i\kappa r},
\end{eqnarray}\newline
where\newline
 $\kappa=\sqrt{2m(E_F+i\omega_n+\frac{i}{2\tau}{\rm sign}\;\omega_n)}$. (We use the value of the root which has the same sign of the imaginary part as $\omega_n$.) Finally we obtain\newline
\begin{eqnarray}
\label{free4}
\sigma^i(r)=2J_{df}S^iT\sum_{\omega_n}G^2(i\omega_n,r).
\end{eqnarray}\newline
Eq. (\ref{free4}) leads to\newline
\begin{eqnarray}
\label{free13}
J_{eff}(R_{ij})=2J_{df}^2T\sum_{\omega_n}G^2(i\omega_n,R_{ij}).
\end{eqnarray}\newline
To take into account spin polarization of the conduction electrons we should modify Eq. (\ref{free13}) to\newline
\begin{eqnarray}
\label{free13b}
J_{eff}(R_{ij})=J_{df}^2T\sum_{\omega_n,\alpha}G_{\alpha}^2(i\omega_n,R_{ij}),
\end{eqnarray}\newline
where\newline
\begin{eqnarray}
G_{\alpha}(i\omega_n,{\bf r})=-\frac{m}{2\pi r}e^{i\kappa_{\alpha} r},
\end{eqnarray}\newline
and $\kappa_{\pm}=\sqrt{2m(E_F\pm J_{df}S \sigma/2+i\omega_n+\frac{i}{2\tau}{\rm sign}\;\omega_n)}$,
where $\sigma =<S^z>/S$ is the  reduced magnetization.
\section{Finite electron band width model}
In this model the dispersion law is given by:\newline
\begin{eqnarray}
\label{free13c}
E_k=\frac{W}{2}\left(1-cos\left( ka \right)\right)
\end{eqnarray}\newline
where a is the lattice constant,W is the conduction bandwidth and k is the wave vector.
The effective exchange in Eq.(\ref{free13b}) is replaced by:\newline
\begin{eqnarray}
\label{free13d}
&&J_{eff}(R_{ij})=-\frac{4z}{W} \left(\frac{V J_{df}}{4\pi^2 N_0 R_{ij}}\right)^2 P \int^\frac{\pi}{a}_0\int^\frac{\pi}{a}_0
[f\left(E_k +J_{df}S \sigma/2\right)+\nonumber\\
&&+f\left(E_k -J_{df}S \sigma/2\right)]
\frac{k sin\left(k R_{ij}\right)k'sin\left(k'R_{ij}\right)}{cos\left(k' a\right)-cos\left(k a\right)}dkdk'
\end{eqnarray}\newline
where $P$ means the Cauchy principal part of the integral over k', z is the degeneracy of the conduction band
(assumed to be 2 in our case), not including the spin. $V$ is the volume of the crystal and $f \left( E \right)$ is Fermi function.
\section{Effective spin Hamiltonian}
The calculations, the results of which are presented below, are based on the effective Hamiltonian\newline
\begin{eqnarray}
H&=&-\frac{1}{2}\sum_{ij} I({\bf R}_{ij}) {\bf S}_i{\bf S}_j-\frac{1}{2}\sum_{ij} J_{eff}({\bf R}_{ij}) S_i^zS^z_j,
\end{eqnarray}\newline
where $J_{eff}(R_{ij})$ is given by Eq.(\ref{free13b}) or by Eq.(\ref{free13d}).
As one can see, for the case of polarized electrons we are getting some anisotropic Heisenberg model for the indirect exchange. However, in the mean field approximation(MFA) only z direction is relevant.
The sum in Eq.(\ref{free13b}) can be calculated only numerically.
Further on we'll present the calculations corresponding to $\tau=\infty$, leaving the analysis of the influence of the finite scattering times for the future publication \cite{burg}.
The emergent magnetic problem is solved
in the mean field approximation. The equation for the reduced magnetization $\sigma$ is \cite{mauger}\newline
\begin{eqnarray}
\sigma=B_S\left[\frac{\sigma S^2\left(I(0)+\sum_jJ_{eff}(R_{ij})\right)+g \mu_BS B_{ext}}{T}\right]\nonumber\\
\label{free14}
\end{eqnarray}\newline
where $B_S$ is the Brillouin function, $g$ is the g factor, $\mu_B$ is Bohr magneton and $B_{ext}$ is the external magnetic field.
The quantity $I(0)$ we chose to fit the $T_C$ of the stoichiometric compound.
Because the average magnetization of localized spins enters into the integral, the calculations should be done selfconsistently.
We are taking in consideration the Eu ions have a structure form of FCC (Face Center Cubic) lattice and the sum over $R_{ij}$ is done according to Table I.
\begin{table}[H]
\begin{tabular}{|c|c|c|}
	\hline
Order &  Number of neighbors    &   Distance between neighbors  \\
	\hline
1 &12 & $\sqrt{1/2} $\\
2 &6 & 1 \\
3 &24& $\sqrt{3/2} $\\
4 &12& $\sqrt{2} $\\
5 &24& $\sqrt{5/2}$ \\
6 &8& $\sqrt{3}$ \\
7 &48& $ \sqrt{7/2}$ \\
8  &6& 2 \\
9 &36& $3/\sqrt{2}$ \\
10 &24&$ \sqrt{5}$ \\
11 &24& $\sqrt{11/2}$ \\
12 &24& $\sqrt{6}$ \\
13 &72&$ \sqrt{13/2}$ \\
	\hline
\end{tabular}
\caption{FCC Number of neighbors and Distance between neighbors.}
\label{free18}
\end{table}
Following Mauger we chose $J_{df}=0.13$ eV for EuO. We take density of states effective mass $m$ being equal to the free electron mass \cite{nagaev}
\section{Results}
\subsection{Free electrons model}
\begin{figure}[H]\includegraphics[height=45mm,width=90mm]{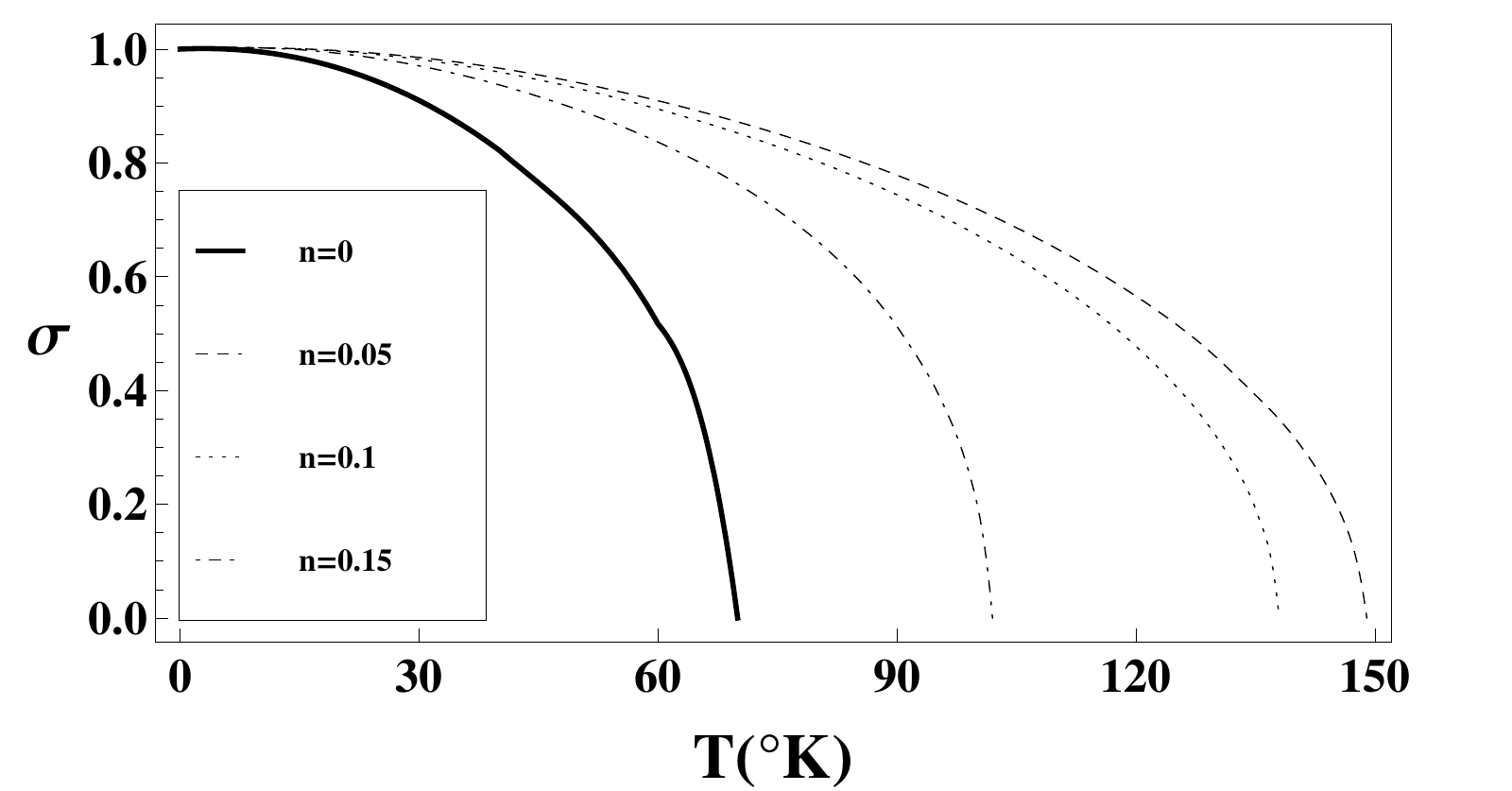}
\caption{The magnetization as a function of temperature ($T$) for $J_{df}=0.13$ eV and for different electron concentrations:
 $n=0$ (Thick line),
 $n=0.05$ (Dashed line),
 $n=0.1$ (Dotted line) and
 $n=0.15$ (DotDashed line).}
\label{free5}
\end{figure}

\begin{figure}[H]\includegraphics[height=45mm,width=90mm]{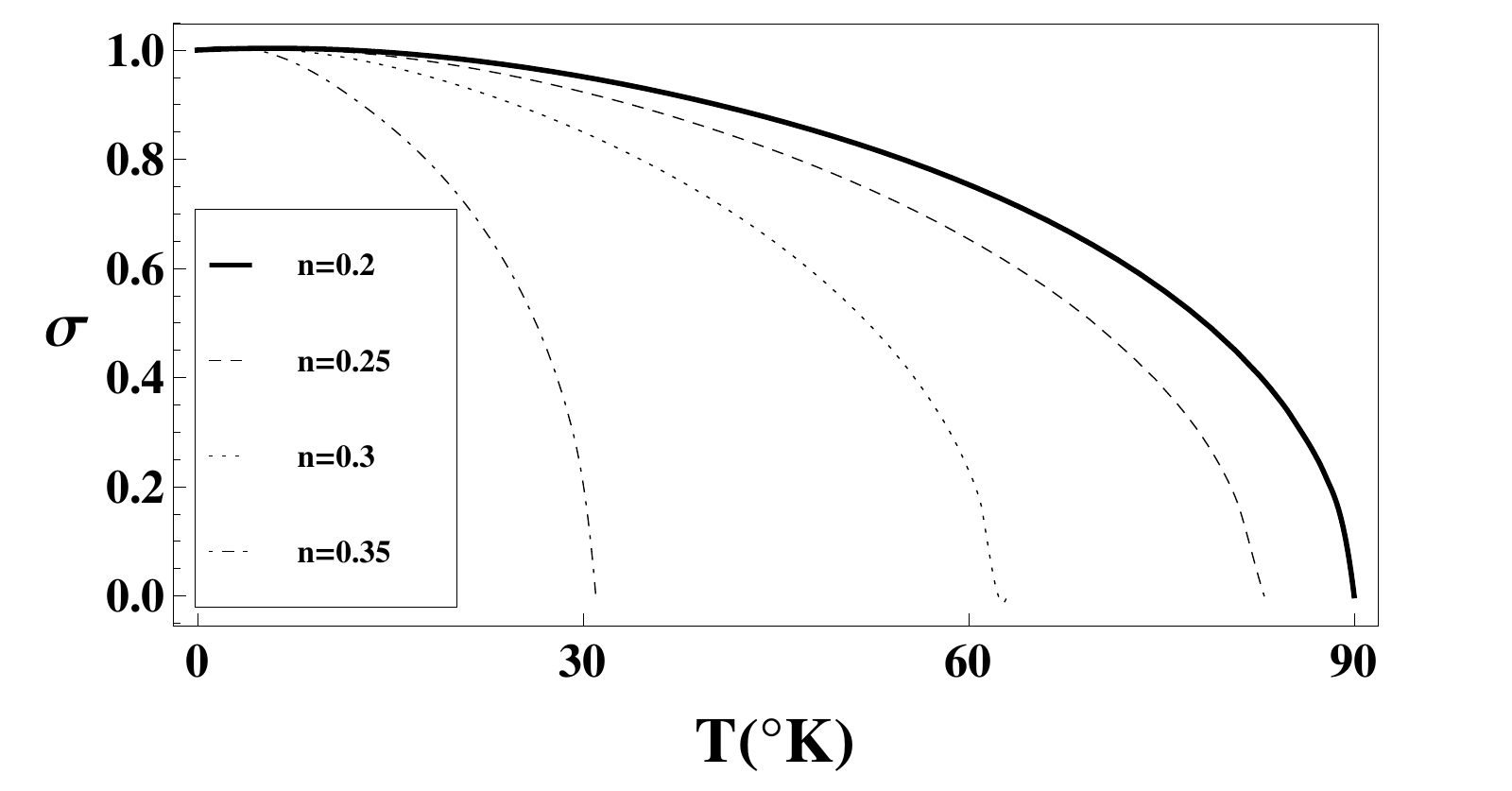}
\caption{The magnetization as a function of temperature ($T$) for $J_{df}=0.13$ eV and for different electron concentrations:
 $n=0.2$ (Thick line),
 $n=0.25$ (Dashed line),
 $n=0.3$ (Dotted line) and
 $n=0.35$ (DotDashed line).}
\label{free15}
\end{figure}

\begin{figure}[H]\includegraphics[height=45mm,width=90mm]{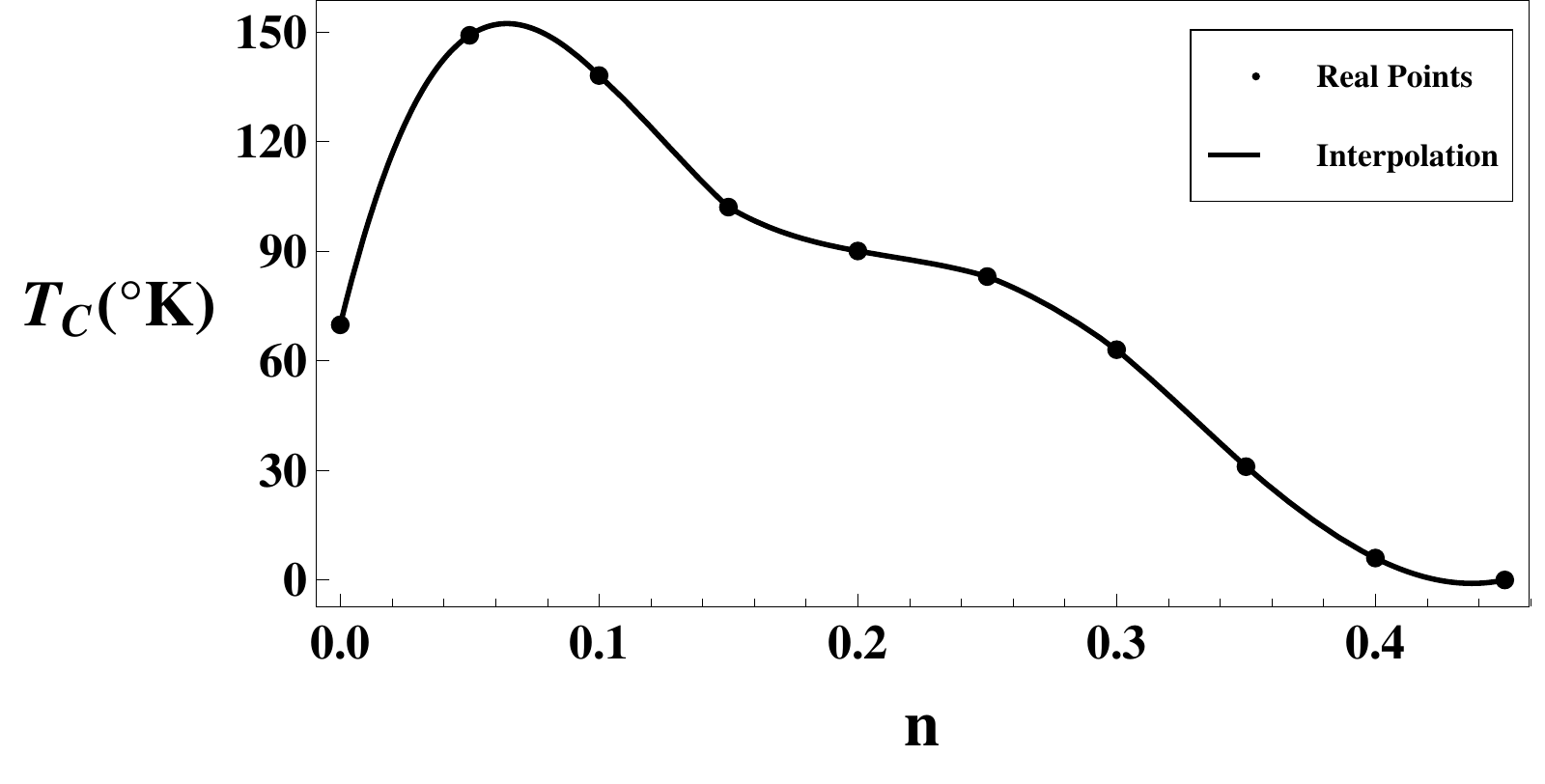}
\caption{Curie temperature ($T_C$) as a function of electron concentration ($n$) for $J_{df}=0.13$ eV where the points represent real points and the line is interpolation.}
\label{free6}
\end{figure}

First, we calculated the magnetization ${\sigma }$ as a function of temperature for different electron concentrations according to Eq.(\ref{free14}) in the absence of external magnetic field. The results are presented in Fig.\ref{free5} and \ref{free15}.
Fig.\ref{free5} and \ref{free15} shows clearly that the magnetization is like Brillouin function even for $n\neq0$. we see also how the indirect exchange ($n\neq0$) changes the critical temperature compared to Curie temperature ($69.8 K$) of stoichiometric compound ($n=0$).
We plotted Curie temperature $T_C$ as a function of electron concentration $n$ on Fig.\ref{free6}.
In Fig.\ref{free6} we see that Curie temperature has maximum value ($149 K$) at electron concentration of $n=0.05$ and return to the temperature of stoichiometric compound($69.8 K$) at electron concentration of $n=0.3$. For $n = 0.45$ the Curie temperature drops to zero, which means that above this concentration the ferromagnetic phase does not exist even at $T=0$.

The reason for the suppression in Curie temperature for large electron concentrations is that as the doping or oxygen vacancies level increases, the wavelength of the effective exchange oscillations ($1/k_F$) becomes shorter and,  hence, increasingly, anti-ferromagnetic, which ultimately suppresses the ferromagnetic transition.

\begin{figure}[H]\includegraphics[height=42mm,width=84mm]{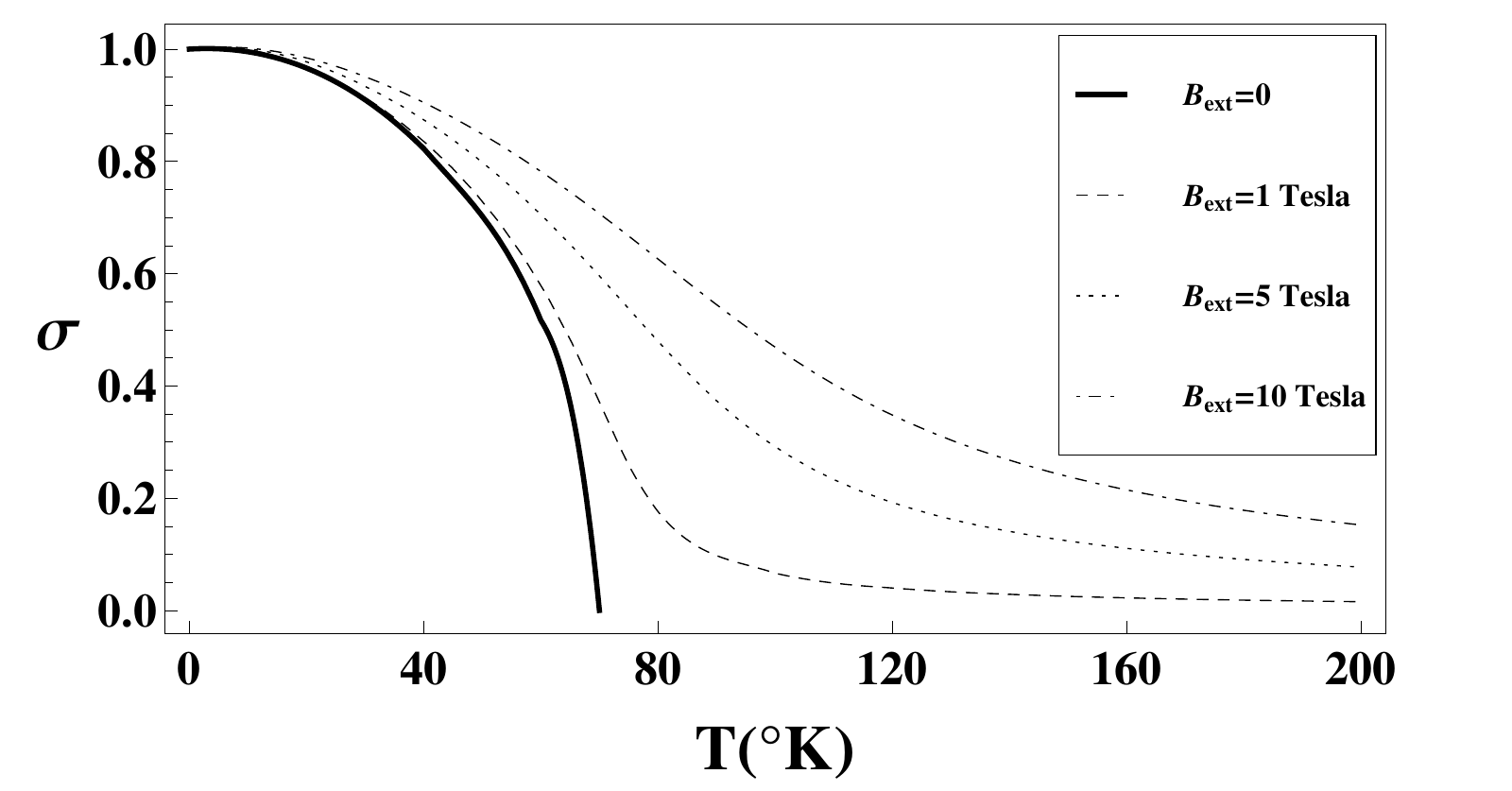}
\caption{The magnetization as a function of temperature ($T$) for $J_{df}=0.13$ eV and $n=0$ and for different external magnetic fields:
$B_{ext}=0$ (Thick line),
$B_{ext}=1$ Tesla (Dashed line),
$B_{ext}=5$ Tesla (Dotted line) and
$B_{ext}=10$ Tesla (DotDashed line).}
\label{free9}
\end{figure}

\begin{figure}[H]\includegraphics[height=45mm,width=90mm]{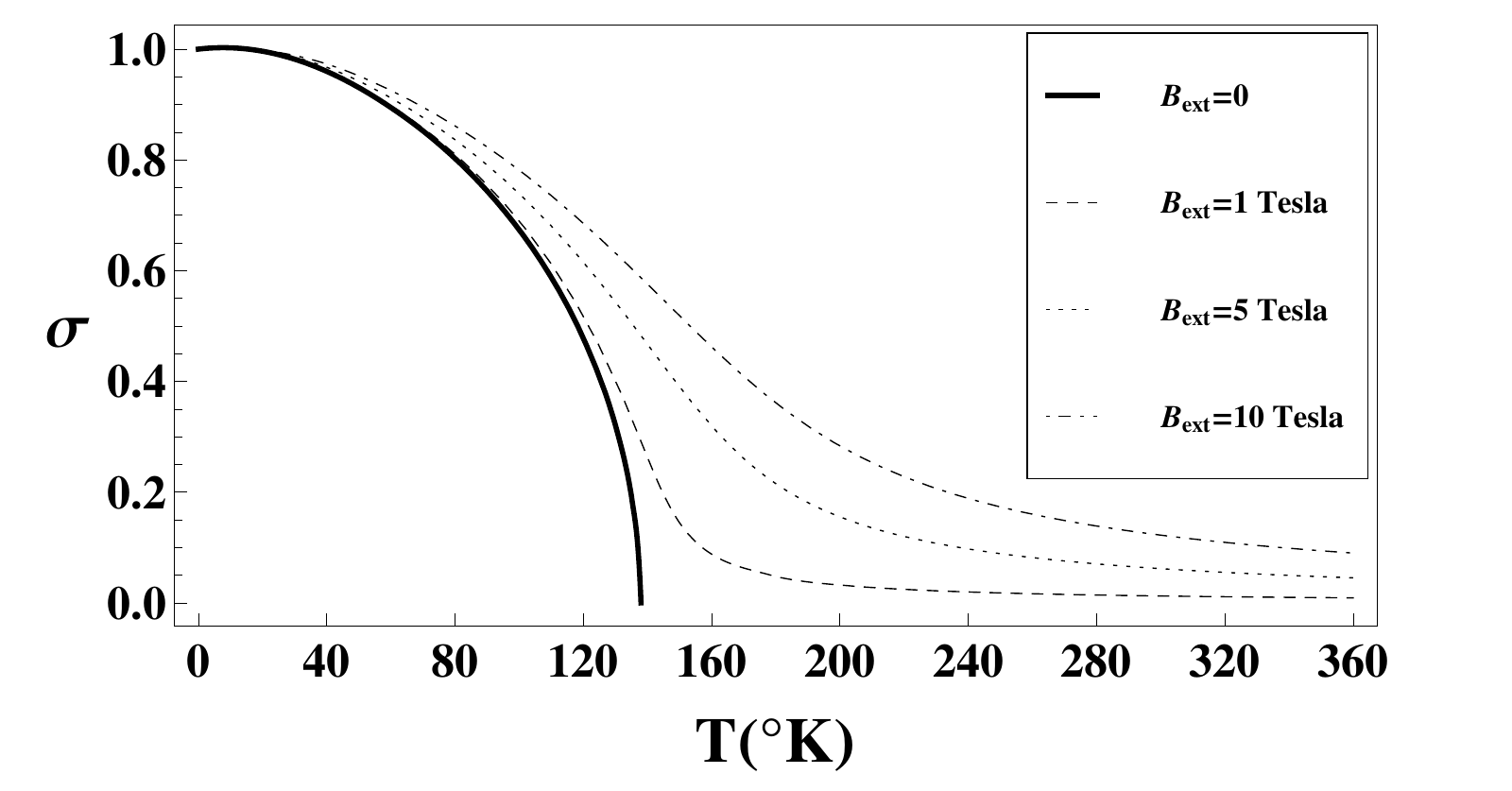}
\caption{The magnetization as a function of temperature ($T$) for $J_{df}=0.13$ eV and $n=0.1$ and for different external magnetic fields:
$B_{ext}=0$ (Thick line),
$B_{ext}=1$ Tesla (Dashed line),
$B_{ext}=5$ Tesla (Dotted line) and
$B_{ext}=10$ Tesla (DotDashed line).}
\label{free10}
\end{figure}

\begin{figure}[H]\includegraphics[height=45mm,width=90mm]{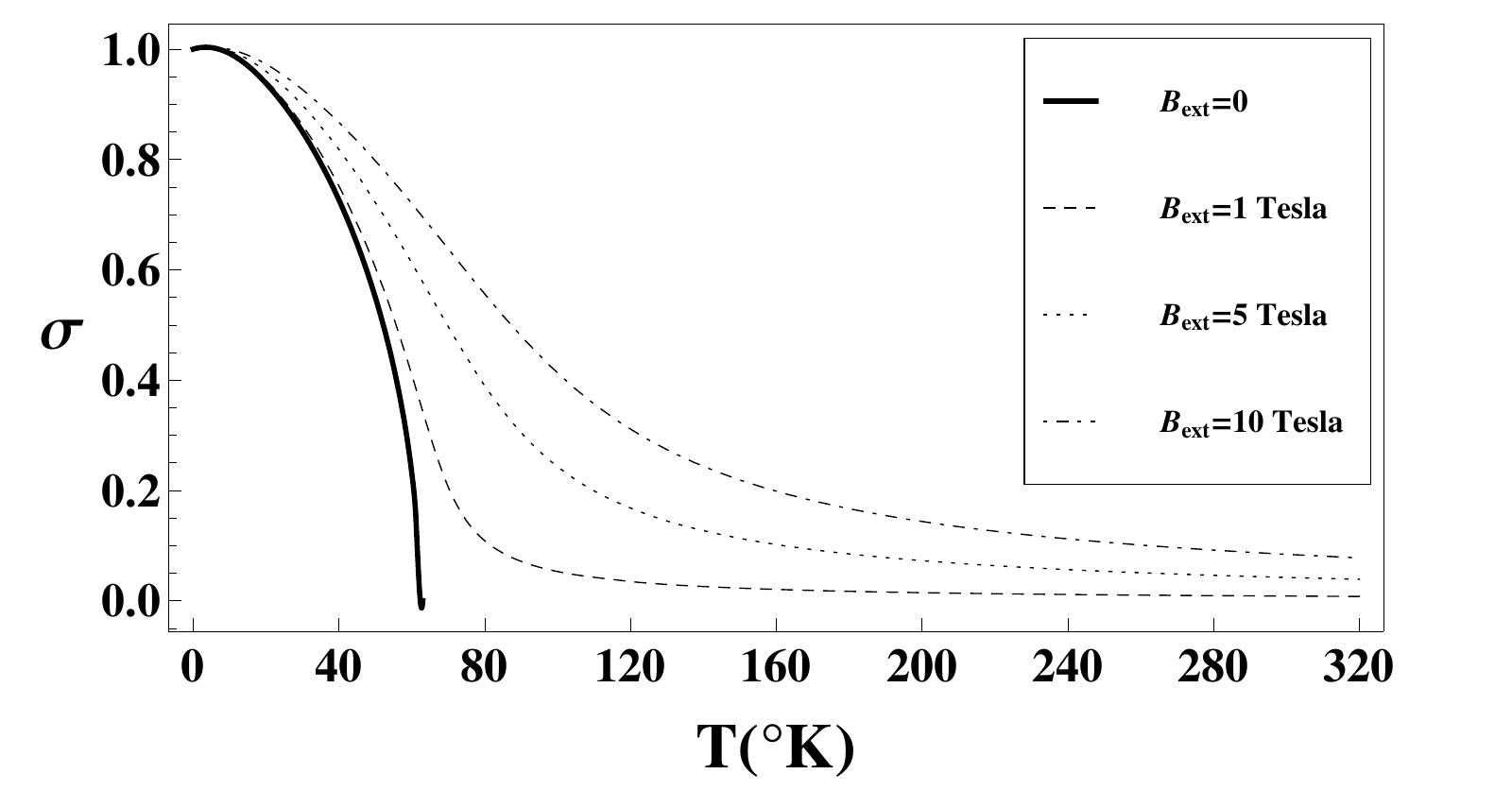}
\caption{The magnetization as a function of temperature ($T$) for $J_{df}=0.13$ eV and $n=0.3$ and for different external magnetic fields:
$B_{ext}=0$ (Thick line),
$B_{ext}=1$ Tesla (Dashed line),
$B_{ext}=5$ Tesla (Dotted line) and
$B_{ext}=10$ Tesla (DotDashed line).}
\label{free11}
\end{figure}

Now we use Eq.(\ref{free14}) to calculate the magnetization ${\sigma }$ as a function of temperature for different electron concentrations and for different external magnetic fields. The results are presented in Fig. \ref{free9}-\ref{free11}.
Fig.\ref{free9}-\ref{free11} shows clearly that the magnetization is not like brillouin function for $n\neq0$ because of the existence of an external magnetic field.
\subsection{Finite electron band width model}
\begin{figure}[H]\includegraphics[height=45mm,width=90mm]{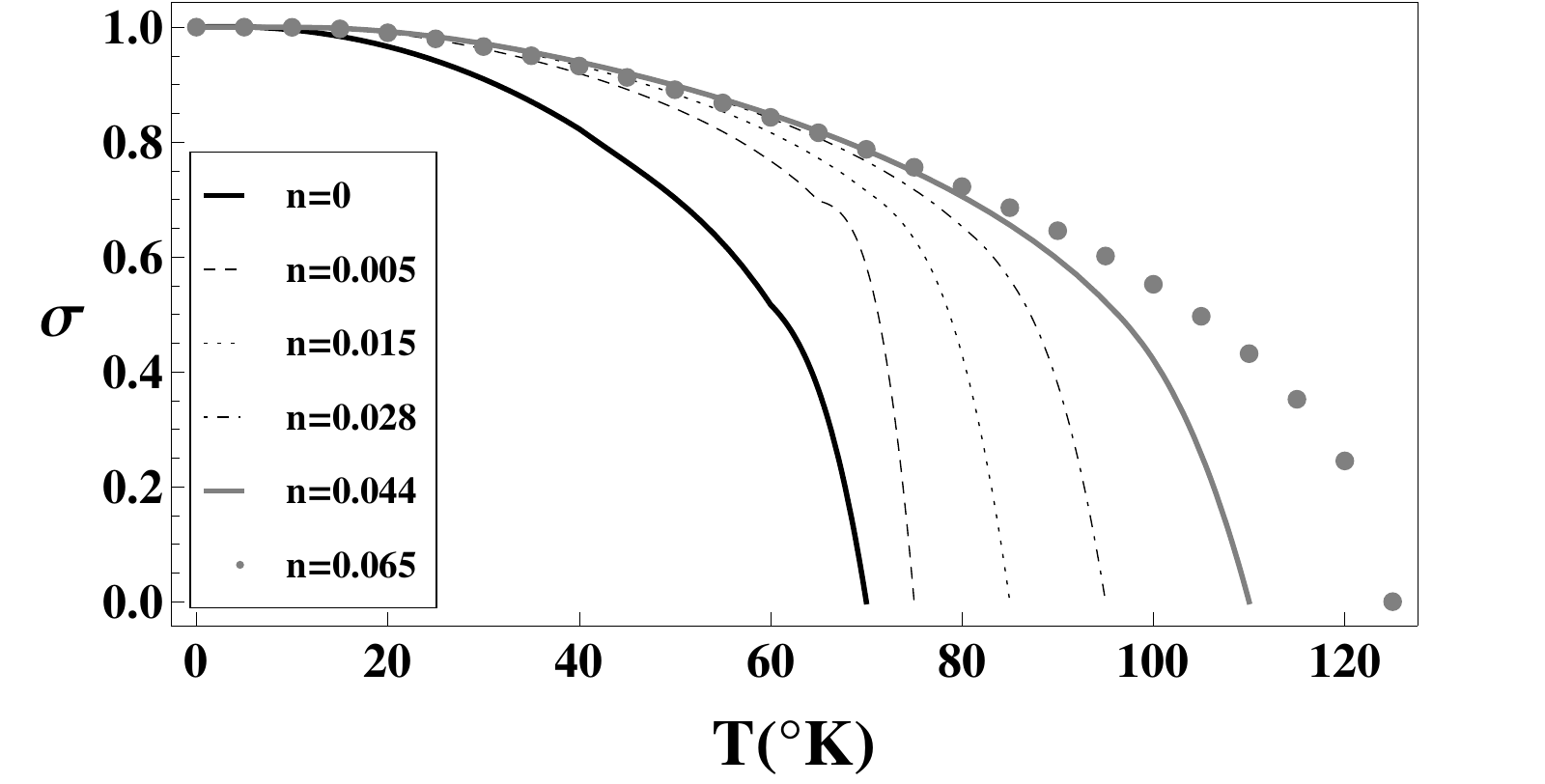}
\caption{The magnetization as  a function of temperature ($T$) according to free electrons model for $J_{df}=0.13$ eV and for different electron concentrations:
 $n=0$ (Black Thick line),
 $n=0.005$ (Black Dashed line),
 $n=0.015$ (Black Dotted line),
 $n=0.028$ (Black DotDashed line),
 $n=0.044$ (Gray Thick line) and
 $n=0.065$ (Gray Dotted line).}
\label{free16}
\end{figure}

\begin{figure}[H]\includegraphics[height=45mm,width=90mm]{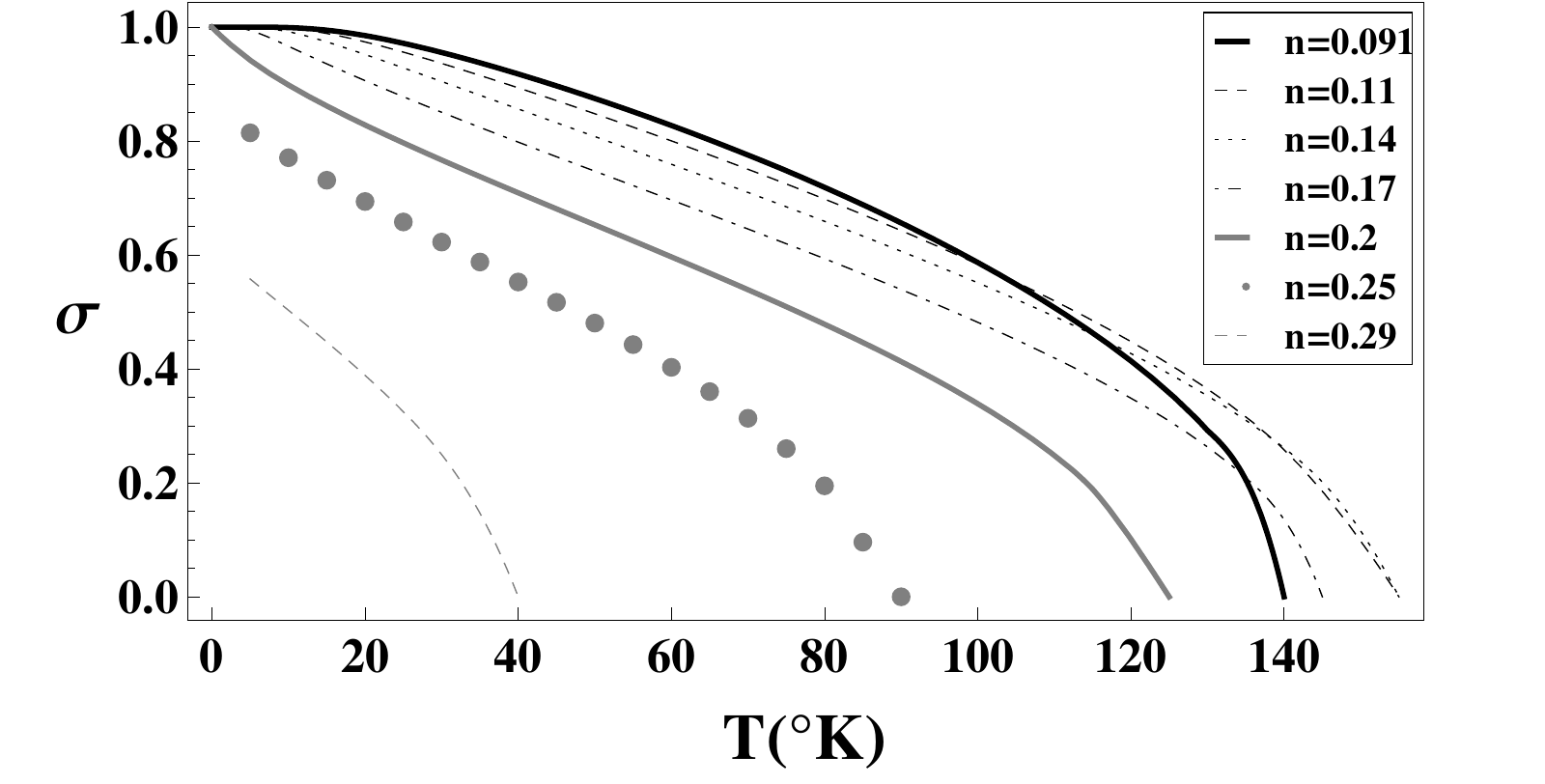}
\caption{The magnetization as a function of temperature ($T$) according to Mauger model for $J_{df}=0.13$ eV and for different electron concentrations:
 $n=0.091$ (Black Thick line),
 $n=0.11$ (Black Dashed line),
 $n=0.14$ (Black Dotted line),
 $n=0.17$ (Black DotDashed line),
 $n=0.2$ (Gray Thick line),
 $n=0.25$ (Gray Dotted line) and
 $n=0.29$ (Gray Dashed line).}
\label{free17}
\end{figure}

\begin{figure}[H]\includegraphics[height=45mm,width=90mm]{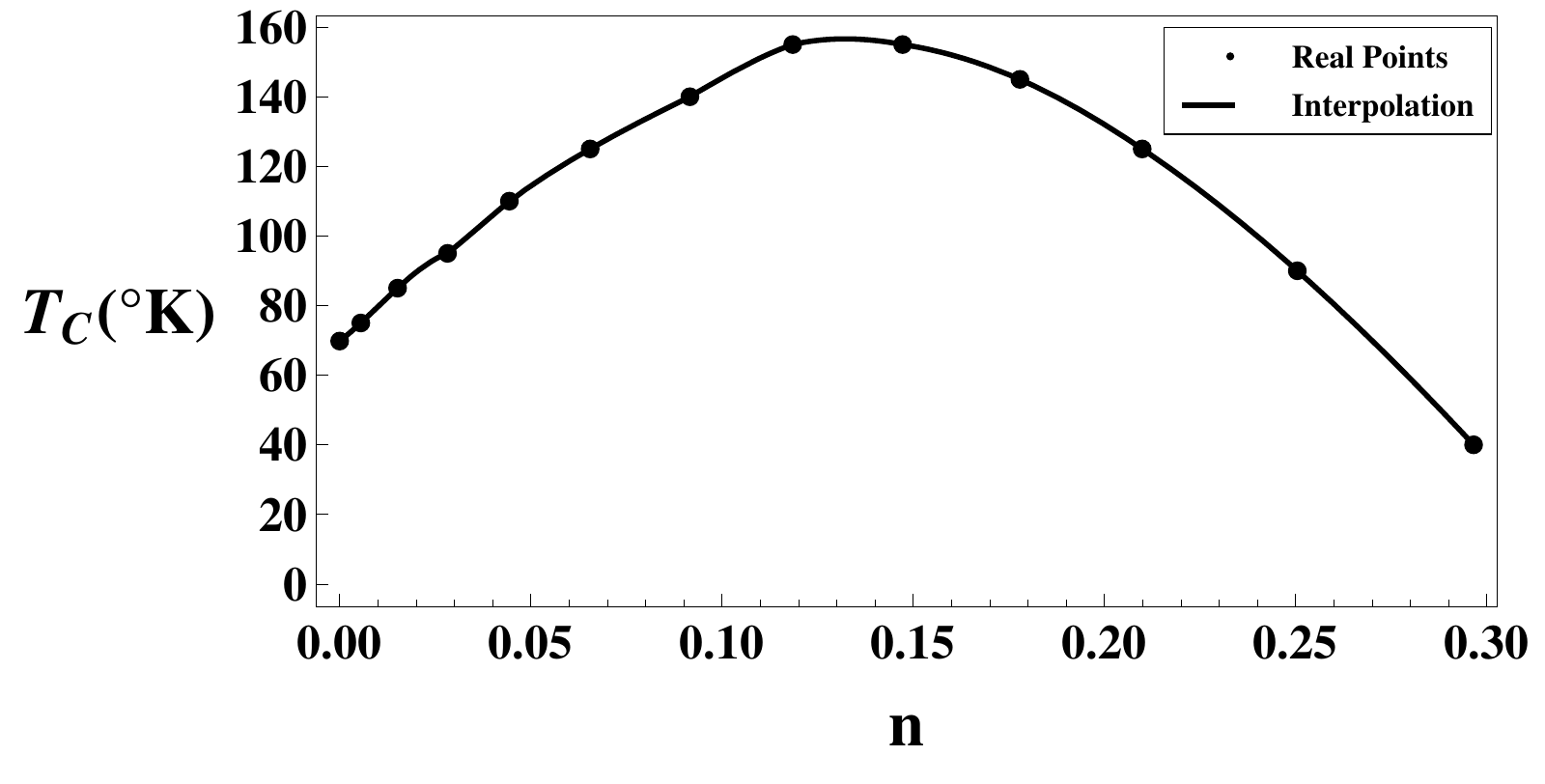}
\caption{Curie temperature ($T_C$) as a function of electron concentration ($n$)- Mauger's result~}
\label{free19}
\end{figure}

First, as before, we calculated the magnetization ${\sigma }$ as a function of temperature for different electron concentrations according to Eq.(\ref{free14}) in the absence of external magnetic field. The results are presented in Fig.\ref{free16} and \ref{free17}.
We plotted Curie temperature $T_C$ as a function of electron concentration $n$ on Fig.\ref{free19}.
In Fig. \ref{free19} we see that Curie temperature has maximum value ($155 K$) around electron concentration of $n=0.1$ and return to the temperature of stoichiometric compound ($69.8 K$) around electron concentration of $n=0.25$. \newline
Like before, the reason for the suppression in Curie temperature for large electron concentrations is that as the doping or oxygen vacancies level increases, the wavelength of the effective exchange oscillations ($1/k_F$) becomes shorter and,  hence, increasingly, anti-ferromagnetic, which ultimately suppresses the ferromagnetic transition.\newline
If we do a comparison between the two models we can see that the concentration dependence of the Curie temperature is similar: the maximum Curie temperature achieved for the optimal doping is almost the same, the concentrations of electrons corresponding to optimal doping are very close. However, for the free electrons model (Fig. \ref{free6}) the decrease of the Curie temperature with the increase of electron concentration is slower than in the finite electron band width model (Fig. \ref{free19}).

\begin{figure}[H]\includegraphics[height=45mm,width=90mm]{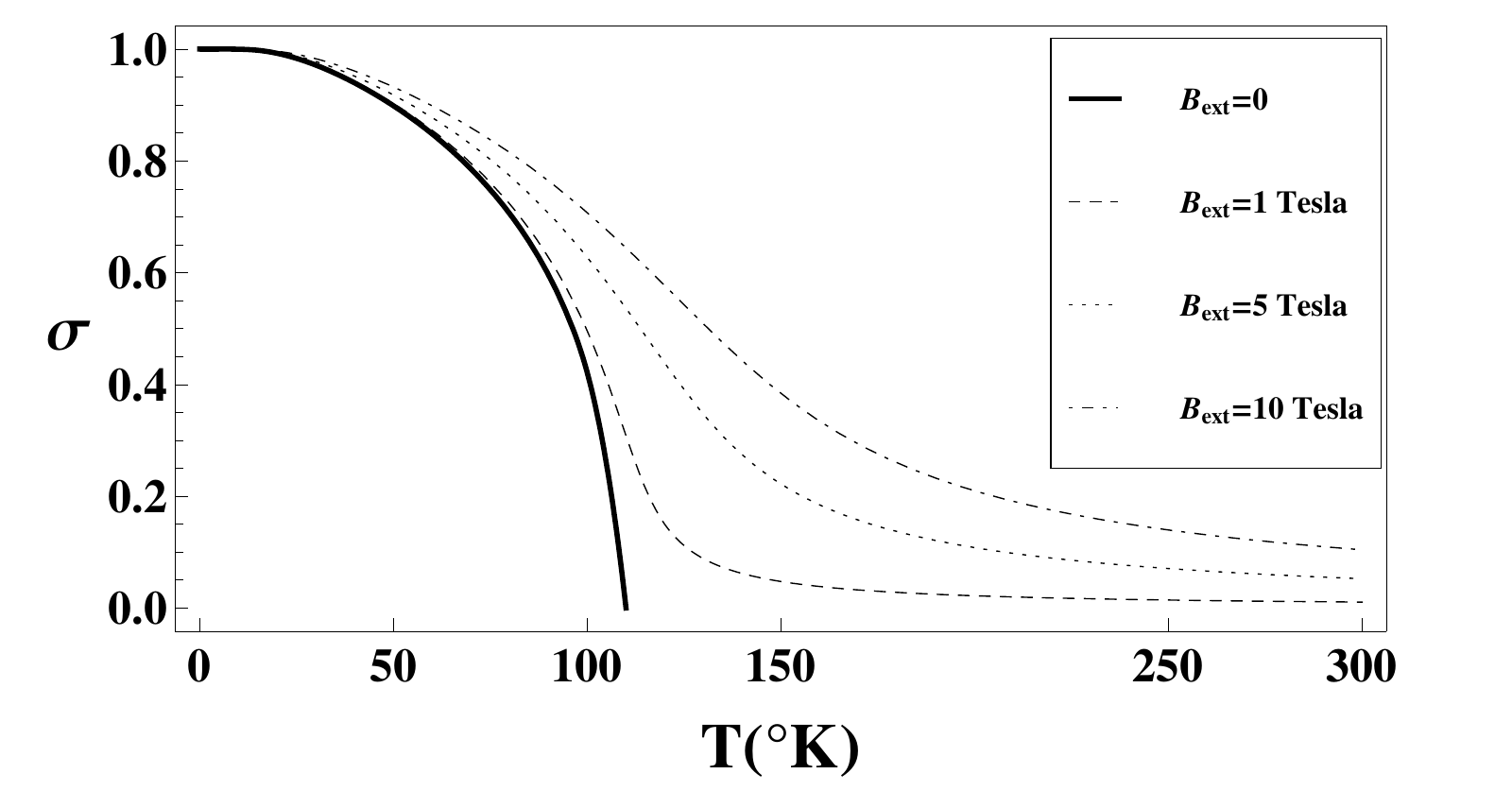}
\caption{The magnetization as a function of temperature ($T$) for $J_{df}=0.13$ eV and $n=0.044$ and for different external magnetic fields:
$B_{ext}=0$ (Thick line),
$B_{ext}=1$ Tesla (Dashed line),
$B_{ext}=5$ Tesla (Dotted line) and
$B_{ext}=10$ Tesla (DotDashed line).}
\label{free20}
\end{figure}

\begin{figure}[H]\includegraphics[height=45mm,width=90mm]{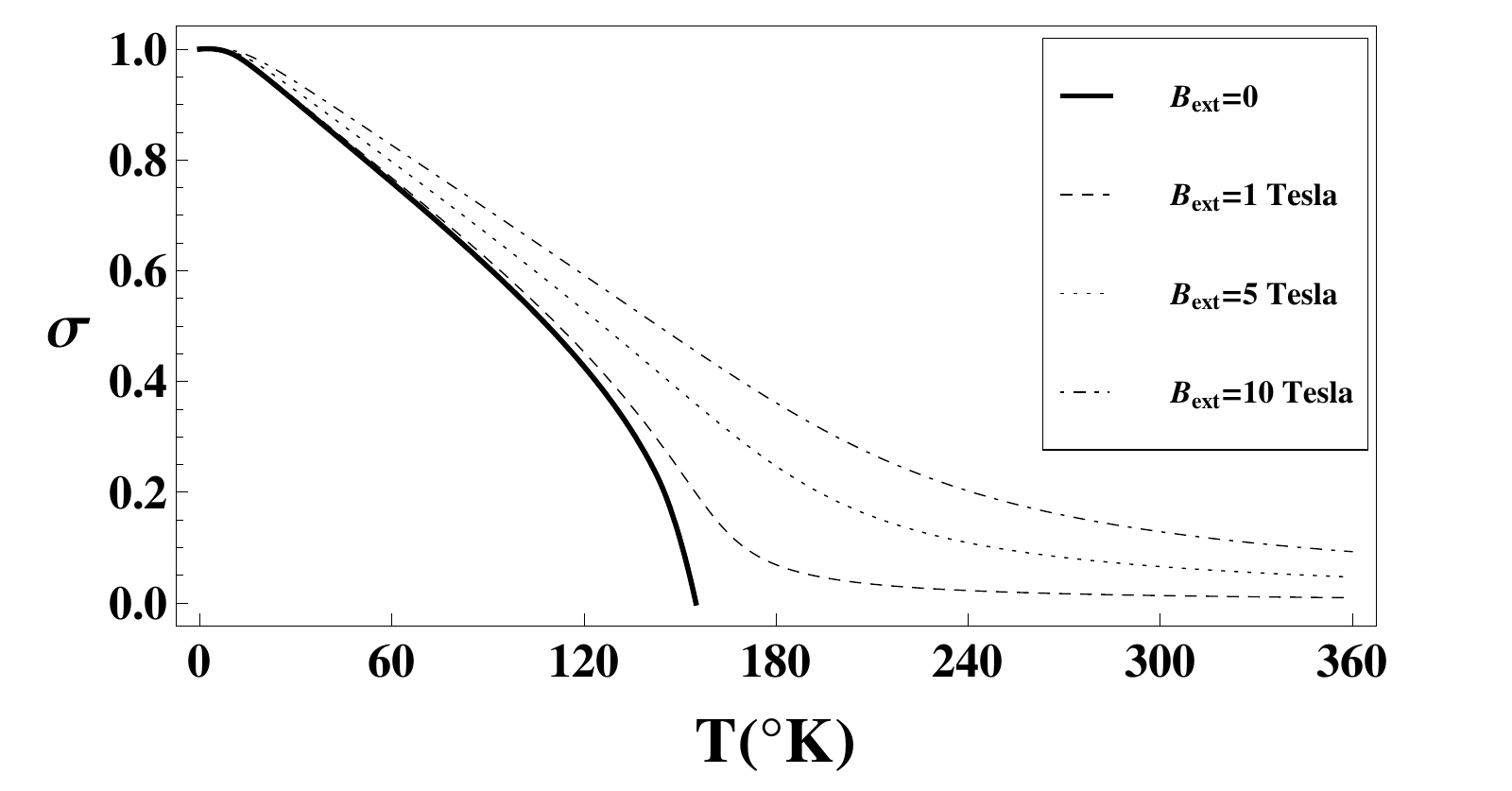}
\caption{The magnetization as a function of temperature ($T$) for $J_{df}=0.13$ eV and $n=0.14$ and for different external magnetic fields:
$B_{ext}=0$ (Thick line),
$B_{ext}=1$ Tesla (Dashed line),
$B_{ext}=5$ Tesla (Dotted line) and
$B_{ext}=10$ Tesla (DotDashed line).}
\label{free21}
\end{figure}

\begin{figure}[H]\includegraphics[height=45mm,width=90mm]{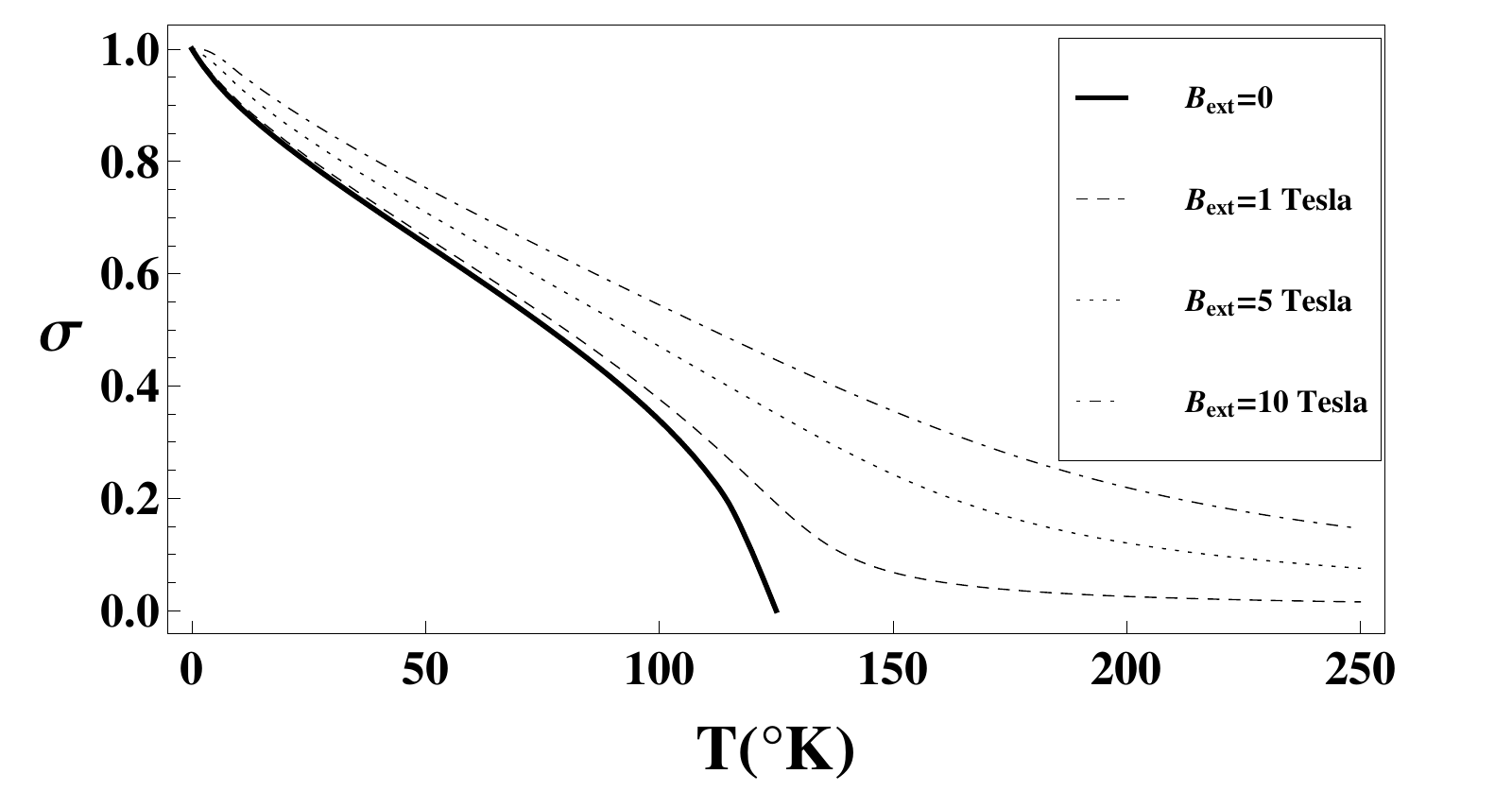}
\caption{The magnetization as a function of temperature ($T$) for $J_{df}=0.13$ eV and $n=0.2$ and for different external magnetic fields:
$B_{ext}=0$ (Thick line),
$B_{ext}=1$ Tesla (Dashed line),
$B_{ext}=5$ Tesla (Dotted line) and
$B_{ext}=10$ Tesla (DotDashed line).}
\label{free22}
\end{figure}

As before, we calculated the magnetization ${\sigma }$ as a function of temperature for different electron concentrations and for different external magnetic fields. The results are presented in Fig. \ref{free20}-\ref{free22}.
\setlength{\parskip}{0pt}
\setlength{\parsep}{0pt}
Again, we can see clearly that the magnetization is not like Brillouin function for $n\neq0$ because of the existence of an external magnetic field.

\section{Conclusions}
We presented the calculations for the influence of the indirect exchange on the magnetic properties of EuO in the framework of Matsubara Green functions technique.
We used different electron dispersion laws: free electrons and finite conduction electron band width.  Qualitatively the results for the concentration dependence of the Curie temperature and temperature dependence of the magnetization were similar, the maximum Curie temperature achieved for the optimal doping is the same, the concentrations of electrons corresponding to optimal doping are very close. However for the free electrons model the decrease of the Curie temperature with the increase of electron concentration is slower than in the finite electron band width model. The temperature dependence of the magnetization in the former case looks closer to the Brillouin function.
However, all experimental results on doped EuO shows a non-like Brillouin magnetization curve and it seems to have a second dome in the magnetization curve. Some authors argue \cite{kroha}\cite{Ott}\cite{Matsumoto},therefor, that the second dome represents a second critical temperature ($T_C$). However, in our calculations we did not see any deviation from Brillouin function for doped EuO in the absence of an external magnetic field .One of the possible explanations for  this contradiction could be the existence of an external magnetic field in experimental systems.\newline
Also, some experiments show values of Curie temperature which are larger than our results ($T_C = 170K$ \cite{Ott} and $T_C = 180K$ \cite{Ahn}). Nevertheless , we emphasize that both models we used ,despite their simplicity, have a good match to most of experimental results of temperature vs. electron concentration curve and the high Curie temperature values we have mentioned, are exceptional .\newline
In this paper, we explained the existence of maximum of $T_C$ by ferromagnetic and anti-ferromagnetic ordering which increases and reduces respectively the indirect exchange. However, some experimental works \cite{mairoser} show the fact that only a small fraction of the introduced dopants act as a donor. These results could provide an alternative explanation the for saturation (or maximum) of $T_C$ which stems from the saturation and decreasing of the concentration of "mobile" charge carriers.\newline
The problem of the influence of doping was recently analyzed in Ref. \cite{savrasov}.
It was shown there  that there are two competing factors.  On one hand, like it was traditionally considered, free carriers  induce the Ruderman-Kittel-Kasuya-Yosida (RKKY) interaction. On the other hand,  since the bottom of the conduction band consists mainly of the majority spin, the doped electron will enter the spin-polarized manifold, and this results in the onset of moment in the 5d band. Both these factors were studied in Ref. \cite{savrasov}
for EuO using the virtual crystal approximation. The second factor was not taken into account
in the main body of our paper (and the first was taken into account in a mode advanced approximation, than virtual crystal).\newline
The comparison of the results of two approaches (free electron model and finite electron band width model) between themselves and with the experiment can give an idea, what features in the observed experimental behavior are robust (the dispersion law independent), and what aren't. Thus it may help to understand the physics of magnetic semiconductors.

\end{document}